\documentclass[aps,prb,reprint,amsmath,amssymb,superscriptaddress,longbibliography]{revtex4-2}
\usepackage[bookmarks=false,linkcolor=NavyBlue,urlcolor=NavyBlue,colorlinks,citecolor=NavyBlue]{hyperref}
\usepackage{graphicx}
\usepackage{amsmath}
\usepackage[svgnames]{xcolor}
\usepackage{enumerate}
\usepackage{comment}

\renewcommand{\vec}[1]{\mathbf{#1}}

\newcommand{\Cornell}{Department of Physics, Cornell University, Ithaca, NY 14853, USA}
\newcommand{\Berkeley}{Department of Physics, University of California, Berkeley, CA 94720, USA}
\newcommand{\EqualContribution}{O.L. and S.B. contributed equally to this work.}

\def \r{{\boldsymbol{r}}}
\def \tn{\textnormal}

\newcommand{\figref}[2]{\hyperref[#1]{\ref{#1}(#2)}}
\newcommand{\figsref}[2]{\hyperref[#1]{\ref{#1}#2}}

\begin{document}

\title{Gaplessness from disorder and quantum geometry in gapped superconductors}
\author{Omri Lesser}
\thanks{\EqualContribution}
\affiliation{\Cornell}

\author{Sagnik Banerjee}
\thanks{\EqualContribution}
\affiliation{\Cornell}

\author{Xuepeng Wang}
\affiliation{\Cornell}

\author{Jaewon Kim}
\affiliation{\Berkeley}

\author{Ehud Altman}
\affiliation{\Berkeley}

\author{Debanjan Chowdhury}
\affiliation{\Cornell}

\begin{abstract}
It is well known that disorder can induce low-energy Andreev bound states in a sign-changing, but fully gapped, superconductor at $\pi-$junctions. Generically, these excitations are localized. Starting from a superconductor with a sign-changing and nodeless order parameter in the clean limit, here we demonstrate a mechanism for increasing the localization length associated with the low-energy Andreev bound states at a fixed disorder strength. We find that the Fubini-Study metric associated with the electronic Bloch wavefunctions controls the localization length and the hybridization between bound states localized at distinct $\pi-$junctions. We present results for the inverse participation ratio, superfluid stiffness, site-resolved and disorder-averaged spectral functions as a function of increasing Fubini-Study metric, which indicate an increased tendency towards delocalization. The low-energy properties resemble those of a dirty nodal superconductor with gapless Bogoliubov excitations. We place these results in the context of recent experiments in moir\'e graphene superconductors. 
\end{abstract}
\maketitle

\section{Introduction} Quantum geometry (QG), while being an old mathematical concept~\cite{provost}, is increasingly emerging as a useful organizing principle for the study of two-dimensional materials~\cite{torma_essay}. Recent theoretical advances have established that the geometric properties of single-electron Bloch wavefunctions~\cite{vanderbilt2018berry} --- beyond the conventional Berry curvature --- are critical for understanding the rich landscape of interacting phases in these systems. Constraints on QG offer a nontrivial mechanism ~\cite{Roy,CRPHYS,ledwith2020fractional,TS20,cano,andrews24,hetenyi_fluctuations_2023} for the formation of exotic phases such as fractional Chern insulators~\cite{TN11,BAB11,XGW11,Sheng2011} and other strongly correlated states in flat bands~\cite{QGreview,carmichael_probing_2025}. The role of QG, which depends on the orbital embedding but is still gauge invariant~\cite{simon_contrasting_2020}, has been particularly illuminated in the context of flat-band superconductivity~\cite{torma2015,Torma16b,Torma17,Torma19,Bernevig19,Rossi19,Huber21,Zhang2021,hofmann_superconductivity_2023,Zhaoyu,zhang_identifying_2025,li_vortex_2025, thumin_strengthening_2024}, a topic that has attracted intense interest following remarkable experimental breakthroughs in layered two-dimensional materials~\cite{Balents2020,Andrei2021,Nuckolls2024}.

\begin{figure}[ht!]
    \centering
    \includegraphics[width=\linewidth]{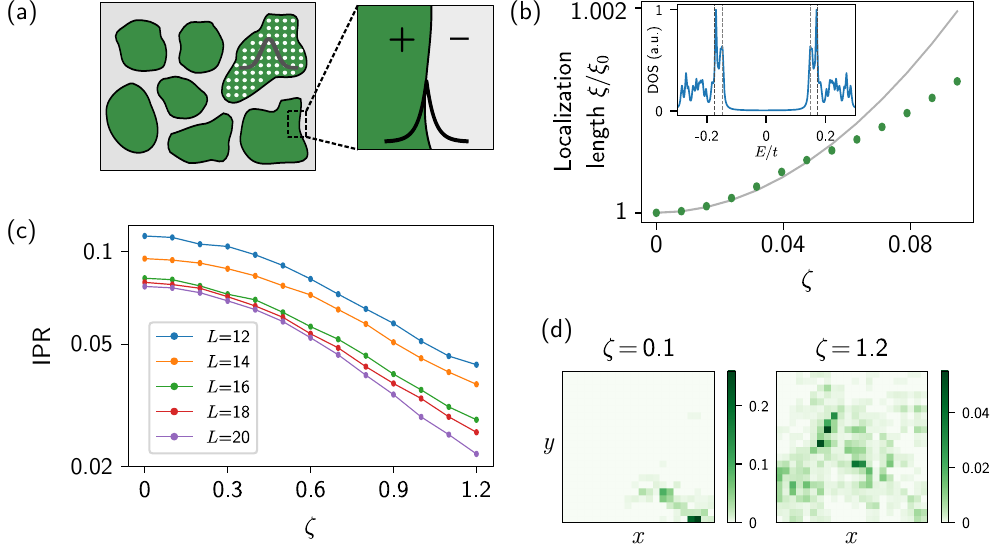}
    \caption{(a)~Schematic illustration of disorder-induced islands (green and gray) with sign-changing gaps; a $\pi-$junction hosts localized Andreev bound states (ABS). Quantum metric controls the minimal spread of the Bloch wavefunctions across many lattice sites (white dots).
    (b)~ABS localization length $\xi$ at a single $\pi-$junction normalized by $\xi_0\sim v^\star_{\rm F}/\Delta$ as a function of $\zeta$; see Eq.~\eqref{eq:H_chiral_band}. The green dots are obtained from a self-consistent mean-field calculation for the model defined in Eqs.~\eqref{model}, \eqref{eq:H_int} on an $18\times18$ lattice, with a domain wall in $V(\vec{r})=V_{0}+V_{d}\Theta(x)$ ($V_{0} = 0.5t$, $V_{d} = 1.6t$);
    gray curve is based on Eq.~\eqref{eq:H_BdG_1D}.
    Inset: DoS in the uniform system, showing extended $s$-wave SC gaps ($\Delta_1\pm\Delta_0$).
    (c)~IPR for the correlated-disordered model ($V_{1}=0.5t,\:V_{2}=2.1t$) averaged over a fixed energy interval $E/L=0.012t$ for a system of size $L\times L$ for different values of $\zeta$, and fixed $\ell=0.1$. 
    (d)~Snapshots of the wavefunctions (spatial probability distributions) at small and large $\zeta$. The other parameters, fixed for all simulations are $t' = 0.18t$, and $U = 0.5t$.}
    \label{fig:general_picture}
\end{figure}

 While the possibility of unconventional superconductivity driven by the combination of strong electronic correlations and QG continues to motivate this rapidly evolving field, the ubiquitous presence of disorder in electronic solids presents an exciting uncharted new frontier. The subtle quantum effects arising from the interplay of quenched disorder, interactions, and QG in two-dimensional superconductors with isolated nearly flat bands remains largely unexplored. Impurities are known to play an important role in correlated high-temperature superconductors~\cite{AlloulRMP}, and disorder has been argued to produce many fascinating new phenomena~\cite{DC15,hirschfeld18,Kivelson21,EA24,bashan,hirschfeld25,bouzerar_robustness_2025,tanaka_symmetry_2012,lau_universal_2022,chan_disorder_2025}. It is well known that disorder can induce subgap bound states in superconductors~\cite{beenakker_random-matrix_1997,sauls_andreev_2018}.
 Flat-band superconductivity exhibits significant robustness against disorder, demonstrated by a quadratic suppression of the critical temperature under weak off-diagonal disorder, contrasting with the linear decrease found in conventional superconductors~\cite{bouzerar_robustness_2025}. While the critical disorder strength required to destroy superconductivity is proportional to the clean system's superfluid weight~\cite{chan_disorder_2025}, other analyses conclude that the disorder-induced suppression of the superfluid weight is universally independent of quantum geometry and band dispersion across diverse $s$-wave models~\cite{lau_universal_2022}.
 One of the central questions we address in this Letter is how the underlying quantum geometry associated with the electronic wavefunctions affects the properties of these impurity-induced subgap states in a superconductor with puddles of sign-changing gap. Specifically, we will be interested in how the combination of spatially inhomogeneous puddles combined with the quantum geometry-induced spatial extent of electronic wavefunctions affect the low-energy thermodynamic and spectral properties of an intrinsically gapped superconductor.

Our starting point will be a two-dimensional interacting electron problem in an inhomogeneous superconductor with domains of {\it sign-changing} order parameters $+\Delta_{1},-\Delta_{2}$; see Fig.~\figref{fig:general_picture}{a}. These domains have a characteristic size $\ell$, and regions of $|\r|\lesssim\ell$ are assumed to be locally homogeneous and have a well developed superconducting gap. In the model we introduce below, such a state is not realized purely by an on-site attractive interaction, $U$,  and requires introduction of a nearest-neighbor attractive interaction $V$. To introduce inhomogeneities, we promote $V(\vec{r})$ to be a random variable that takes values $V_{1}, V_{2}$ with equal probability, and with exponentially decaying spatial correlations with correlation length $\ell$, $\left\langle V(\vec{r})V(\vec{r'}) \right\rangle\propto\exp(-|\vec{r}-\vec{r}'|^2/2\ell^2)$, In such a setting, as we will demonstrate below, when $|V_{1}-V_{2}|$ is large enough, we obtain sign-changing domains with characteristic size $\ell$. Clearly, the sign change of the order parameter between islands (a $\pi-$junction) induces Andreev bound states~\cite{beenakker_random-matrix_1997}. The localization length of the Andreev bound states is known to be controlled by the intrinsic length scale of the superconductor, i.e., the coherence length $\xi$. In a superconductor whose ground state is well described by the Bardeen--Cooper--Schrieffer (BCS)  many-body wavefunction, this length scale is $\xi_{0}\sim v^\star_{\rm F}/\Delta$, where $v^\star_{\rm F}$ is the renormalized Fermi velocity. However, the above description makes no reference to any intrinsic length scale associated with the electronic states that contribute to pairing. Since the quantum metric effectively controls the minimal spread associated with the Wannier wavefunctions~\cite{MarzariVanderbilt}, we expect it to alter $\xi$. Recent theoretical considerations, that do not include the effects of disorder, have already suggested that the coherence length in a uniform superconductor is enhanced due to a quantum-geometric contribution~\cite{hu_anomalous_2025,chen_ginzburg-landau_2024,sun_flat-band_2024}.

\section{Low-energy description} The basic building block for studying the network of superconducting puddles and their associated low-energy excitations is a single $\pi-$junction. Consider a low-energy theory focusing on the spatial dependence of the gap structure in the direction transverse to the domain wall; thereby the problem can be treated within an effectively one-dimensional Bogoliubov--de Gennes (BdG) framework. 
We include a dispersion term $iv^\star_{\rm F}\partial_{x}$, that contains renormalization due to the interactions. The effect of the quantum geometry is modeled as a $\vec{k}$-dependent pair function~\cite{hu_anomalous_2025}, $\Delta(\vec{k})\approx\Delta_0(1-\zeta^{2}a^{2}\vec{k}^2)$, where $\zeta$ accounts for the quantum geometry and $a$ is a microscopic (lattice) length scale. Here, all elements in the quantum geometric tensor $\mathcal{G}_{ij}(\mathbf{k})$ are proportional to $\zeta^{2}a^2$; as we will show in the context of the specific model we introduce below.
This form of $\Delta(\mathbf{k})$ originates from writing the projected pairing interaction $\sum_{\vec{k},\vec{q}}[\Delta(\vec{k})\Lambda(\vec{k},\vec{k}+\vec{q}) c_{-\vec{k}}c_{\vec{k}+\vec{q}} + \rm{H.c.}]$, where $\vec{q}$ is the momentum transfer and $\Lambda(\vec{k},\vec{k}+\vec{q}) = u^{*}(\vec{k}+\vec{q})u(\vec{k})$ is the overlap between the two Bloch states, $u(\vec{k})$, at $\vec{k}$ and $\vec{k}+\vec{q}$. We note that $\vec{k}$ is expanded near the Fermi momentum $k_{\rm F}$, whereas the projected form factor is expanded near $\vec{q}=0$.
The resulting BdG Hamiltonian is then 
\begin{equation}\label{eq:H_BdG_1D}
    H_{\rm 1D} = iv^\star_{\rm F}\partial_{x}\hat{\tau}_{z} + \Delta_0\left(x\right)\left(1+\zeta^{2}a^{2}\partial_{x}^{2}\right)\hat{\tau}_{x},
\end{equation}
where the Pauli matrices $\hat{\tau}_{j}$ act in particle-hole space, and $\Delta_0(x)=\Delta_{0}{\rm sign}(x)$ encodes the $\pi-$junction structure. Assuming the $x$ dependence of the solution only comes from a decaying envelope of the form $\exp(-|x|/\xi)$, we find that for zero-energy bound states,
\begin{equation}\label{eq:effective_xi}
    \xi=\frac{1}{2}\left(\xi_{0} + \sqrt{\xi_{0}^{2}+4\zeta^{2}a^{2}}\right),    
\end{equation}
where $\xi_{0}=v_{\rm F}^{\star}/\Delta_{0}$, which is the intrinsic geometry-independent estimate for the localization length.
As we demonstrate in Fig.~\figref{fig:general_picture}{b}, this form (gray lines) approximately matches a self-consistent numerical calculation for a microscopic model at small $\zeta$, whose details will be presented shortly.
Corrections for states with nonzero energy $E$ are suppressed by ${\cal O}(E^2/\Delta^2)$, and therefore we neglect them for the purpose of this discussion. As we suspected, the coherence length is enhanced by the quantum geometric contribution. This sets the stage for our discussion of a networks of disorder-induced $\pi-$junctions, where the hybridization between nearby localized states can be controlled by the quantum geometry, raising the question of how the low-energy thermodynamic and spectroscopic properties evolve with increasing $\zeta$.

\section{Microscopic model} Our interest lies in the thermodynamic and spectral features associated with the asymptotic low-energy subgap excitations ($0\lesssim E\ll\Delta$). We expect the disorder-induced many-body density of energy levels to affect the thermodynamic and spectral response, and their wavefunctions to control the (de-)localized characteristics, and relatedly the transport observables. It is useful to place these results in the context of a concrete microscopic model with independently tunable bandwidth (i.e., Fermi velocity) and quantum geometry. Moir\'e materials, where many of the theoretical ingredients that are central to our study are relevant, host a panoply of competing orders, triggered by a complex interplay of electron-electron and electron-phonon interactions. In what follows, we refrain from making direct connections with microscopic models of contemporary moir\'e materials, and focus instead on a simplified model Hamiltonian to illustrate a variety of proof-of-concept questions.  Another advantage of this model is that it has been solved previously in the {\it clean} limit using a combination of controlled analytical and numerically exact methods ~\cite{hofmann_heuristic_2022,hofmann_superconductivity_2023}. In this manuscript, we extend the model to incorporate the effects of disorder in the interaction strength. 

The two-dimensional model is defined on the square lattice with $H=\sum_{\vec{k},l,s}c^{\dagger}_{\vec{k},l,s}[H(\vec{k})-\mu \hat{\tau}_{0}\hat{\sigma}_{0}]c_{\vec{k},l,s} +  H_{\text{int}}$, where $c^{\dagger}_{\vec{k},l,s}$ is a creation operator for an electron with momentum $\vec{k}$, orbital $l = 1,2$, and spin $s = \uparrow, \downarrow$. The Hamiltonian takes the form
\begin{subequations}
\begin{eqnarray}
H(\vec{k}) &=& H_{\rm{flat}}(\vec{k}) + H_{\rm{disp}}(\vec{k})\label{model},\\
\label{eq:H_chiral_band}
    H_{\rm{flat}}\left(\vec{k}\right) &=& -t\left(\hat\lambda_{x}\sin\alpha_{\vec{k}}+\hat\sigma_{z}\hat\lambda_{y}\cos\alpha_{\vec{k}}\right),\\
    H_{\text{disp}}\left(\vec{k}\right) &=& -t^{\prime}\left[\cos(k_{x}a) + \cos(k_{y}a)\right]\hat\lambda_0\hat\sigma_0,
\end{eqnarray}
\end{subequations}
where the Pauli matrices $\hat\sigma_{j},\hat\lambda_{j}$ act in spin and orbital space, respectively, and $   \alpha_{\vec{k}}=\zeta\left[\cos(k_{x}a)+\cos(k_{y}a)\right]$. Notice that the bands obtained from $H_{\rm{flat}}(\vec{k})$ are perfectly flat with energies $\varepsilon_{\vec{k}}=\pm t$ and a band gap $E_{\rm{band}}=2t$ (which is then reduced due to the dispersive term $t'<t$), and are topologically trivial regardless of the value of $\zeta$. Their quantum metric, when integrated over the Brillouin zone, is $\zeta^{2}a^{2}/2$~\cite{hofmann_superconductivity_2023}. Note that even though the minimal spatial extent of the maximally (exponentially) localized Wannier functions for the flat-band of $H_{\rm{flat}}\left(\vec{k}\right)$ is controlled by $\zeta$, the band-dispersion remains perfectly flat as a result of a non-trivial quantum interference effect. To include an independently tunable bandwidth, which can also originate from interactions, we have included the term $H_{\rm{disp}}(\vec{k})$ that induces a finite Fermi velocity and, correspondingly, a finite coherence length in the superconductor, $\xi_0$, even when $\zeta\rightarrow0$. 

Given our interest in superconducting phases at low temperatures, we explicitly introduce local and frequency-independent {\it attractive} interactions, that include an onsite ($U$) and a nearest-neighbor ($V$) component:
\begin{equation}\label{eq:H_int}
    H_{\text{int}}=-U\sum_{\vec{r},l}\hat{n}_{\vec{r},l}^{2}-\sum_{\left\langle \vec{r},\vec{r'}\right\rangle }V\left(\vec{r}\right) \hat{n}_{\vec{r}}\hat{n}_{\vec{r}'},
\end{equation}
where $\hat{n}_{\vec{r},l}=\sum_{s}c^{\dagger}_{\vec{r},l,s}c_{\vec{r},l,s}$ is the occupation at site $\vec{r}$ and orbital $l$, and $\hat{n}_{\vec{r}}=\sum_{l}\hat{n}_{\vec{r},l}$. We will assume disorder in the nearest-neighbor interactions, with a uniform onsite component. In the remainder of this study, we will set $4t',~U,~V \lesssim 2t$ and work in units where $t=1$. We will also choose a filling such that the ``remote" band with $\varepsilon_{\vec{k}}=+t$ associated with $H_{\text{flat}}(\vec{k})$ is unoccupied. Based on quantum Monte Carlo computations, the ground state for $t'=V=0$ is known to be a robust {\it fully gapped} $s$-wave superconductor at all (partial) fillings of the lower flat band~\cite{hofmann_superconductivity_2023}. While an attractive nearest-neighbor interaction can favor phase separation~\cite{phasesep}, we exclude this possibility here and focus only on the superconducting ground state.
The above model is time-reversal symmetric, and in what follows we assume that the ground-state does not break time-reversal symmetry spontaneously.
Within our low-temperature self-consistent mean-field computations, the nearest-neighbor interaction favors an extended $s$-wave pairing solution which can coexist with the onsite $s$-wave component. In the clean limit, this yields a gap function, $\Delta(\vec{k})=\Delta_0+\Delta_1[\cos(k_xa) + \cos(k_ya)]$, where $\Delta_1/\Delta_0$ is controlled by $V/U$. In the translationally invariant limit (i.e., non-random couplings), the self-consistent superconducting solution is fully gapped, without any (accidental) nodal excitations; see the inset of Fig.~\figref{fig:general_picture}{b}. Note that the nearest-neighbor attraction is crucial for the sign-changing profile of the pair wavefunction and the associated $\pi-$junctions with the inclusion of disorder. %\textcolor{red}{In addition in the clean limit, both onsite and nearest-neighbor interactions are necessary for realizing a nodeless and sign-changing gap function $\Delta(\mathbf{k})$.}%

Before solving the problem with disorder, let us revisit the properties of the bound states at a single $\pi-$junction [Eq.~\eqref{eq:H_BdG_1D}], but realized within the above microscopic model by introducing a domain wall in the pairing interaction oriented along the $\hat{\vec{x}}$ direction, $V(\vec{r})=V_{0}+V_{d}\Theta(x)$ $(V_{0}, V_{d} > 0)$, where $\Theta(x)$ is the Heaviside step function, and uniform $U$. A fully self-consistent numerical solution for the pairing fields and the energy spectrum leads to localized Andreev bound states at the junction, whose localization length is plotted as a function of $\zeta$ in Fig.~\figref{fig:general_picture}{b}. The numerical results agree with the quadrature form of Eq.~\eqref{eq:effective_xi} for small $\zeta$, where the bare coherence length arises from $t'$. Deviations from the analytical form, especially with increasingly larger $\zeta$, are expected due to the states not being exactly at zero energy and the approximation 
vis-\`a-vis dependence of the form-factors on $\zeta$.

\section{Quantum geometry and Andreev bound states} Having established the quantum geometric contribution to the localization length of Andreev bound states at a single junction, we now proceed to study an array of $\pi-$junctions in a disordered superconductor. 
As noted in the introduction, we promote the nearest-neighbor attraction $V(\vec{r})$ to a random variable that takes values $V_{1},V_{2}>0$ with equal probability, and with exponentially decaying spatial correlations set by a lengthscale, $\ell$, while treating the onsite interaction to be uniform with magnitude $U$. 
%To illustrate the scenario of interest, we consider the following form of disorder [Eq.~\eqref{eq:H_int}]: $V(\vec{r})$ takes on the values , $\left\langle V(\vec{r})V(\vec{r'}) \right\rangle\propto\exp(-|\vec{r}-\vec{r}'|^2/2\ell^2)$, with $\ell$ the disorder correlation length. 
In such a setting, when $|V_{1}-V_{2}|$ is large enough, we obtain $\pi-$junctions in the self-consistent pairing fields. Crucially, the origin of the $\pi-$junctions is tied to a Cooper-pair scattering involving states with a sign-changing gap function~\cite{Ambegaokar}. 
We note that the model being studied here belongs to symmetry class CI~\cite{altland_nonstandard_1997}. In spite of the original time-reversal symmetry squaring to $-1$, when restricting to the spin-singlet sector within the BdG formalism, the effective time-reversal symmetry squares to $+1$ whereas the particle-hole symmetry squares to $-1$. Wavefunctions in class CI are always localized in two dimensions~\cite{ma_localized_1985,lee_localized_1993,evers_anderson_2008}. The states we examine below are therefore in principle localized, but we will see that their localization length can be enhanced with increasing Fubini-Study metric.

We now examine the influence of quantum geometry on the Andreev bound states trapped at $\pi-$junctions by varying $\zeta$. We obtain the low-energy many-body spectrum and the associated eigenfunctions for individual disorder configurations via the self-consistent mean-field setup, and focus primarily on disorder-averaged quantities. For diagnosing the localization properties, we compute the inverse participation ratio (IPR), defined for a normalized wavefunction $\psi(\vec{r})$ as ${\rm IPR} = \sum_{\vec{r}} |\psi(\vec{r})|^4$; see Fig.~\figref{fig:general_picture}{c}. The IPR$\lesssim 1$ when the eigenstates are localized on a single lattice site, and exhibits a weak dependence on the system size. On the other hand, for states that are spread over an $O(1)$ fraction of the lattice sites, the IPR shows a decreasing trend with increasing system size; see Fig.~\figref{fig:general_picture}{d}~\cite{kim_fractionalization_2023}. However, as a matter of principle, since the states are fundamentally localized, the IPR eventually saturates to a constant with further increase in system size.

For fixed $\ell$, the bound-state eigenstates' localization is determined by two compounding factors. The first is the band dispersion, quantified by $t^{\prime}$, which sets the effective BCS coherence length $\xi_{0}$. The second factor, and the focus of our study, is the quantum geometric contribution controlled by $\zeta$. In Fig.~\figref{fig:general_picture}{c} we show the IPR averaged over a fixed energy interval of $E/L=0.012t$ (that lies well below the superconducting gap in the clean limit) and over 50 statistically independent disorder configurations. The IPR results do not depend sensitively on either of these numbers, indicating their convergence.
Based on a system-size scaling, we indeed observe an enhanced {\it tendency }towards delocalization even while the states remain localized. From snapshots of the low-energy wavefunctions within a typical disorder configuration, these results are consistent with our expectation of increasing delocalization over a larger fraction of the sites across the system with increasing $\zeta$; see Fig.~\figref{fig:general_picture}{d}.
In our particular model, we have noticed that increasing $\ell$ reduces the local pairing gap, thereby increasing $\xi_0$, and enhancing the localization length starting at smaller values of $\zeta$. We have also studied the energy dependence of the IPR, which exhibits an interesting structure at small $\zeta$.

\begin{figure}
    \centering
    \includegraphics[width=\linewidth]{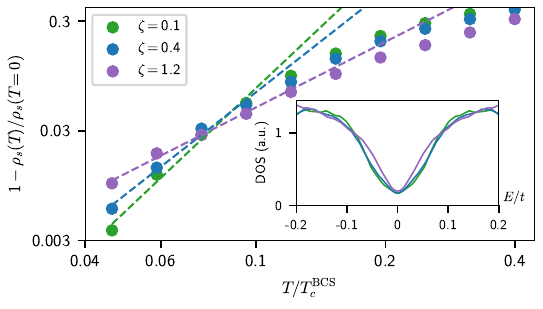}
    \caption{Superfluid stiffness $\rho_s$ as a function of temperature $T$, for three different value of $\zeta$, calculated self-consistently on an $18\times18$ lattice. For all values of $\zeta$, we find a low-temperature power-law behavior of the stiffness indicating gapless superconductivity. The dashed lines are power-law fits, with the exponents varying between $3.7$ at $\zeta = 0.1$, to $2$ at $\zeta = 1.2$, at low temperatures.
    Inset:~DoS for the three $\zeta$ values, showing the low-energy gapless excitations. Other parameters are $t' = 0.18t, U=0.5t, V_{1}=0.5t, V_{2}=2.1t$, and  disorder correlation length $\ell=0.1$.
    }
    \label{fig:stiffness}
\end{figure}

\section{Superfluid stiffness} To examine how the low energy excitations affect the thermodynamic properties of direct relevance to superconductivity in our disordered setup, we study the temperature-dependent superfluid stiffness $\rho_{s}(T)$. For a superconductor with a full spectral gap, the deviation of $\rho_{s}(T)$ from its $T=0$ value is exponentially suppressed at low temperatures. The network of $\pi-$junctions can lead to a non-trivial suppression of the stiffness with increasing temperature \cite{KivelsonSpivak}. In our example, given the gaplessness associated with the disorder-induced states, we anticipate a power law suppression, with the exponent determined by the low-energy (subgap) density of states (DoS). 

We obtain $\rho_s(T)$ by applying a phase twist $\mathbf{q}= q_{x} \hat{x}$ to the pair fields, and calculating the second derivative of the free energy, $\rho_{s}= \frac{1}{L^2} \partial^{2}{\cal F}/\partial q_{x}^{2} |_{q_{x} \rightarrow0}$, where $L$ is the linear system size. Note that the phase twist (or equivalently, a probe vector potential) couples to the current operators that arise from both $H_{\text{disp}}$ and $H_{\text{flat}}$, respectively. The resulting stiffness is plotted in Fig.~\ref{fig:stiffness} for three different values of the $\zeta$ ($T_{c}^{\rm BCS}$ is the temperature at which the order parameter onsets in the disorder-averaged theory, ignoring any Berezinskii-Kosterlitz-Thouless physics that is relevant close to the true superconducting transition \cite{hofmann_superconductivity_2023}). In all cases, we observe an approximate power-law decay of the stiffness at low temperatures, consistent with our picture of gapless superconductivity. The power-law depends on both $\zeta$ and $\ell$ at the lowest temperatures, as indicated by the low-energy difference in the DoS \cite{goldenfeld} (inset of Fig.~\ref{fig:stiffness}; we note a variation in the exponent between $3.7$ at $\zeta = 0.1$, and $2.0$ at $\zeta = 1.2$. The results for $\rho_s(T)$ are similar with increasing $\ell$. Notice that the qualitative behavior of $\rho_s(T)$ remains unchanged over the entire range of $\zeta$ studied here.

\section{Spectral function} 
\begin{figure}
    \centering
    \includegraphics[width=\linewidth]{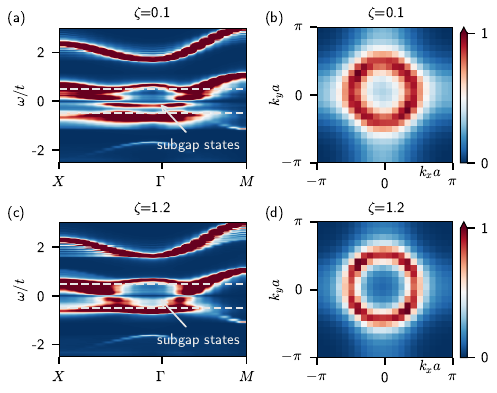}
    \caption{Bogoliubov spectral function ${\cal A}(\vec{k},\omega)$ evaluated at $\zeta=1.2$ on a 22$~\times~$22 lattice (in arbitrary units), for $\zeta=0.1$ and $\zeta=1.2$. (a), (c)~Line cuts of ${\cal A}(\vec{k},\omega)$ along high-symmetry lines in the Brillouin zone. Below the clean pairing gap [dashed lines; see the inset of Fig.~\figref{fig:general_picture}{b}], the localized gapless excitations with enhanced localization length nearly mimic an impurity-band like excitation spectrum. (b), (d)~Momentum-resolved spectral function at zero energy ${\cal A}(\vec{k},\omega=0)$, which suggests that the disorder-induced subgap excitations form a ``Fermi surface". Other parameters are $t' = 0.18t, U=0.5t, V_{1}=0.5t, V_{2}=2.1t$, and $\ell=0.1$.}
    \label{fig:spectral_function}
\end{figure}

To further elucidate the nature of the subgap states, we reconstruct in Fig.~\ref{fig:spectral_function} the momentum-resolved Bogoliubov spectral function ${\cal A}(\vec{k},\omega)$, defined as
\begin{equation}\label{eq:spectral}
    {\cal A}(\vec{k},\omega) \equiv \tn{Im}\sum_{\vec{r}\vec{r'}l;n}\left[\frac{\bar{u}_{n;\vec{r}l}u_{n;\vec{r'}l}}{\omega - E_n + i\eta} + \frac{\bar{v}_{n;\vec{r}l}v_{n;\vec{r'}l}}{\omega + E_n + i\eta} \right]e^{i\vec{k}\cdot(\vec{r}-\vec{r'})},
\end{equation}
where $\psi_n=[u_{n;\vec{r}l},v_{n;\vec{r}l}]^{\tn{T}}$ is the BdG eigenvector associated with energy $E_n$, and the index $n$ sums over the filled states. At small $\zeta$, the spatially localized subgap states below the clean [$V=2.1t$, see Fig.~\figref{fig:general_picture}{b}-inset] pairing gap (dashed lines in Fig.~\ref{fig:spectral_function}) behave like isolated impurity states. A suppressed hybridization due to the small impurity wavefunction overlaps (controlled by $\zeta$) leads to randomly distributed localized energy levels within the ``bulk" gap [Fig.~\figref{fig:spectral_function}{a}]. With increasing $\zeta$, these disorder-induced subgap states mimic an impurity ``band" due to their enhanced localization length, while remaining localized below the original gaps $\omega=\pm\Delta$ [Fig.~\figref{fig:spectral_function}{c}]. We find that after disorder-averaging, these low-energy sub-gap excitations form a ``Bogoliubov Fermi surface" along the original Fermi electronic Fermi surface [Fig.~\figref{fig:spectral_function}{b} and Fig.~\figref{fig:spectral_function}{d}] at both small and large $\zeta$. These low-lying states contribute to the thermodynamical properties in a way that restores the $C_4$ statistical symmetry, which is also consistent with the ``gapless" nature reflected in $\rho_S(T)$ in Fig.~\ref{fig:stiffness}. However, we note that with increasing $\zeta$, the resulting Fermi surface becomes much sharper.
At larger disorder correlation length $\ell$, we find similar results but as argued previously, the increased tendency towards an enhanced localization length onsets at smaller $\zeta$. 
The spatially resolved density of states shows that indeed the system may seem ``gapped" deep within individual domains, and thus its gapless nature arises solely from the Andreev bound states.

\section{Outlook} In this work, we have used a combination of explicit numerical computations on model Hamiltonians and analytical BdG analysis to study the effect of quantum geometry and long-wavelength inhomogeneities on the Andreev bound states in a nominally gapped unconventional superconductor. The low-energy many-body excitations mimic the behavior of dirty ``nodal" Bogoliubov quasiparticles, whose mobility is controlled by quantum geometry. While we have relied primarily on a self-consistent mean-field analysis for our numerical analysis, performing numerically exact quantum Monte-Carlo computations including the effects of disorder remains an exciting future direction. It would also be useful to analyze the frequency resolved optical conductivity in future work, as it will likely reveal an interesting metallic-like response tied to the delocalized subgap excitations coexisting with the superfluid response at zero frequency~\cite{armitage19}. Relatedly, analyzing the thermal conductivity can possibly help isolate the contribution to transport from intrinsically mobile nodal excitations vs. localized disorder-induced sub-gap excitations~\cite{taillefer}.
It is also worth mentioning that for systems in the superconducting symmetry classes D and DIII, e.g., a nodeless triplet $p$-wave superconductor, a true delocalization transition is allowed~\cite{evers_anderson_2008}, suggesting the possibility of driving a delocalization transition of Andreev bound states in such systems.

There is increasing experimental evidence for unconventional superconductivity~\cite{Oh2021unconventional,Kim2022,park_simultaneous_2025} and ``intrinsic" nodal quasiparticles in twisted bilayer and trilayer graphene based on recent measurements of $\rho_s(T)$~\cite{tanaka_superfluid_2025,banerjee_superfluid_2025}. At the same time, there is also direct experimental evidence for a variety of sources of inhomogeneity in stacked graphene-based heterostructures~\cite{alden,Kerelsky2019,Xie2019stm,Choi2019,Uri2020}. While our work has made no direct connections to the microscopic models of either of these two materials, it raises an interesting question of how to experimentally discern a clean nodal superconductor from a nominally gapped (but unconventional) superconductor with localized bound states induced by disorder, whose localization length can be greatly enhanced by quantum geometric effects. It would be interesting to also analyze theoretically the non-linear Meissner response associated with the latter, in light of the recent experiments~\cite{banerjee_superfluid_2025}. 
Extending our results to settings with topological (Chern) bands appearing as time-reversed partners~\cite{AV19,wang_intertwined,kim_theory_2025,wang_spin-polaron_2025}, including in other symmetry classes, would also be of interest.
Finally, given the inherent theoretical challenges associated with describing the strong-coupling regimes of moir\'e superconductivity, an alternative route is to focus on general constraints on the diamagnetic response based on optical sum-rules~\cite{DMDC1,DMDC2,MR21,JFMV24}, but now incorporating the effects of disorder explicitly~\cite{Kivelson82,TKN25}. 

\section*{Acknowledgments} We thank H. Steinberg, S. Ilani, S. Kivelson and A. Vishwanath for interesting discussions. D.~C. thanks E. Berg and J. Hofmann for related collaborations. O.~L. is supported by a Bethe-KIC postdoctoral fellowship at Cornell University. D.~C. is supported in part by a NSF CAREER grant (DMR-2237522), and a Sloan Research Fellowship.

%%%%%%%%%%%%%%%%%%%%%%%%%%%%%%%%%%%%%%%%%%%%%%%%%%
%%%% APPENDIX %%%%%%%%%%%%%%%%%%%%%%%
%%%%%%%%%%%%%%%%%%%%%%%%%%%%%%%%%%%%%%%%%%%%%%%%%%
\appendix

\section{Localization length in a single junction}
Here we describe the low-energy theory of a single $\pi-$junction in a band with quantum geometry. As introduced in the main text, the effect of quantum geometry on the pair potential can be described \cite{hu_anomalous_2025} as $\Delta(\vec{k})\approx\Delta_0(1-\zeta^{2}a^{2}\vec{k}^2)$. This leads to the low-energy Hamiltonian 
\begin{equation}\label{SMeq:H_1D}
    H_{1\rm D}=\begin{pmatrix}iv_{\rm F}^{\star}\partial_{x} & \Delta\left(x\right)\left(1+\zeta^{2}a^{2}\partial_{x}^{2}\right)\\
\Delta\left(x\right)\left(1+\zeta^{2}a^{2}\partial_{x}^{2}\right) & -iv_{\rm F}^{\star}\partial_{x}
\end{pmatrix},
\end{equation}
which is Eq.~\eqref{eq:H_BdG_1D} of the main text. We stress that $v_{\rm F}^{\star}$ is impacted by both the intrinsic Fermi velocity of the band, and the bandwidth renormalized by the interactions. To describe a $\pi$ junction, we take $\Delta\left(x\right)=\Delta_{0}\text{sign}\left(x\right)$. 

In order to solve this eigenvalue equation, we treat each side of the junction separately. Anticipating subgap Andreev states, we take the ansatz \begin{equation}
    \begin{pmatrix}\psi_{e}\left(x\right)\\
\psi_{h}\left(x\right)
\end{pmatrix}=\begin{pmatrix}\chi_{e}\\
\chi_{h}
\end{pmatrix}e^{-x/\xi},
\end{equation}
where $\xi$ is the effective localization length. Substituting this into Eq.~\eqref{SMeq:H_1D}, we set the determinant of $E-H$ to zero in order to find a nontrivial solution, resulting in \begin{equation}
    E^{2}+\frac{v_{\rm F}^{\star2}}{\xi^{2}}-\Delta_{0}^{2}\left(1-\frac{\zeta^{2}a^{2}}{\xi^{2}}\right)^{2}=0.
\end{equation}
At this point we define $\epsilon=E/\Delta_{0}$ as the dimensionless energy and $\xi_{0}=v_{\rm F}^{\star}/\Delta_{0}$ as the bare coherence length. The equation becomes \begin{equation}
    \epsilon^{2}+\left(\frac{\xi_{0}}{\xi}\right)^{2}-\left[1-\left(\frac{\zeta a}{\xi}\right)^{2}\right]^{2}=0.
\end{equation}
At $\epsilon=0$, the physical solution of this equation is 
\begin{equation}
    \xi=\frac{1}{2}\left(\xi_{0} + \sqrt{\xi_{0}^{2}+4\zeta^{2}a^{2}}\right),    
\end{equation}
which is Eq.~\eqref{eq:effective_xi} of the main text. Stitching the solutions at $x>0$ and $x<0$ yields the full bound-state wavefunction. Exact expressions may be obtained also for $\epsilon\neq0$, but they are cumbersome; the leading correction to the effective coherence length comes at order $\epsilon^{2}$ so it is suppressed at low energies.

\section{Details of numerical computations}
We employ a self-consistent mean-field decomposition in the pairing channel, for the Hamiltonian with attractive onsite and nearest neighbor interactions [Eq.~\eqref{eq:H_int} of the main text]. To address the effects of disorder, we transform the non-interacting part of the Hamiltonian~\eqref{eq:H_chiral_band} to real space:
\begin{equation}\label{SMeq:real_space_H_nonint}
\begin{aligned}
     \mathcal{H}_{0} &= \sum_{i,j,l,s} h_{1}(i,j,\zeta) \hat{c}_{i,l,s}^{\dagger} (\lambda_{x}\sigma_{0})\hat{c}_{j,l,s} + h_{2}(i,j,\zeta) \hat{c}_{i,l,s}^{\dagger} (\lambda_{y}\sigma_{z})\hat{c}_{j,l,s} \\
     &- t' \sum_{\langle i,j \rangle, l,s} \hat{c}_{i,l,s}^{\dagger} \hat{c}_{j,l,s} - \mu \sum_{i,l,s} \hat{c}_{i,l,s}^{\dagger} \hat{c}_{i,l,s} + \text{H.c},
\end{aligned}
\end{equation}
with $\zeta$-dependent hoppings $h_{1}$ and $h_{2}$ resulting in the flat band part of the dispersion. These are given by
\begin{subequations}\label{SMeq:zeta-hoppings}
\begin{align}
h_{1}(i,j,\zeta)
&= -\frac{t}{2}\, i^{x_{j}+y_{j}-x_{i}-y_{i}-1}
\Big[
    J_{x_{j}-x_{i}}(\zeta)\,
    J_{y_{j}-y_{i}}(\zeta)
\nonumber\\
&\qquad\qquad\qquad
 - J_{x_{j}-x_{i}}(-\zeta)\,
   J_{y_{j}-y_{i}}(-\zeta)
\Big],
\\[1ex]
h_{2}(i,j,\zeta)
&= -\frac{t}{2}\, i^{x_{j}+y_{j}-x_{i}-y_{i}}
\Big[
    J_{x_{j}-x_{i}}(\zeta)\,
    J_{y_{j}-y_{i}}(\zeta)
\nonumber\\
&\qquad\qquad\qquad
 + J_{x_{j}-x_{i}}(-\zeta)\,
   J_{y_{j}-y_{i}}(-\zeta)
\Big].
\end{align}
\end{subequations}
Here $J_{n}(\zeta)$, $n \in \mathbb{Z}$ denote the Bessel functions of the first kind. We made use of the Schläfli's integral representation of the Bessel functions $J_{n}(\zeta) = \frac{1}{2\pi i^{n}} \int_{0}^{2\pi}d\phi\:e^{i\zeta\:\text{cos}\:\phi- in \phi} $ while transforming $\mathcal{H}_{0}$ from momentum space to real space. In all our numerical simulations, we truncate the flat-band hoppings $h_{1}, h_{2}$ after $|i-j|=3$ in both $x$ and $y$ directions, for every lattice site $i$, which introduces a small correction to the bandwidth of order $\zeta^{7}$ for $\zeta\rightarrow0$.

We perform the mean-field decomposition of $\mathcal{H}_{\text{int}}$ for both onsite and nearest-neighbor $s$-wave pairings, given by the mean-field Hamiltonians:
\begin{eqnarray}
    H_{\text{MF}}^{\text{onsite}} &=& \sum_{i,l,l'}\Delta_{l,l'}(\mathbf{r}_{i})\hat{c}_{i,l,\uparrow}^{\dagger}\hat{c}_{i,l',\downarrow}^{\dagger} + \text{H.c}, \label{SMeq:on_site_MF}\\
    H_{\text{MF}}^{\text{nn}} &=& \sum_{\langle i,j \rangle, l,l'}\Delta_{l,l'}(\mathbf{r}_{i},\mathbf{r}_{j})\hat{c}_{i,l,\uparrow}^{\dagger}\hat{c}_{j,l',\downarrow}^{\dagger} + \text{H.c}. \label{SMeq:nn_MF}
\end{eqnarray}
The pairings are diagonal in the orbital channels. The onsite and nearest-neighbor pairings are given by
\begin{align}
\Delta_{l,l'}(\mathbf{r}_{i})
&= U(\mathbf{r}_{i})\,\delta_{l,l'}\,
\langle \hat{c}_{i,l,\uparrow}\hat{c}_{i,l',\downarrow} \rangle
\label{SMeq:pair_onsite}
\\[1ex]
\Delta_{l,l'}(\mathbf{r}_{i},\mathbf{r}_{j})
&= \frac{V(\mathbf{r}_{i})}{2}\,\delta_{l,l'}
\Big[
    \langle \hat{c}_{i,l,\uparrow}\hat{c}_{j,l',\downarrow} \rangle
\nonumber\\
&\qquad\qquad\qquad
 - \langle \hat{c}_{i,l,\downarrow}\hat{c}_{j,l',\uparrow} \rangle
\Big]
\label{SMeq:pair_extended}.
\end{align}
The self-consistent iterations operate at a fixed electron filling $\langle n \rangle =  0.3$ of the lower band. The chemical potential $\mu$ is determined from $\frac{1}{L^2}\sum_{\mathbf{r}_{i},l,s}n_{i,l,s} = \langle n \rangle$ at every iteration, where $L$ is the linear system size.

\subsection{Calculation of localization length for a single $\pi$-junction}\label{SM:single_junction}
For the mean-field computation, we use a domain-wall form of the nearest-neighbor pairing interaction $V(\mathbf{r}) = V_{0}+V_{d}\Theta(x)$ with $V_{0} = 0.5t$ and $V_{d}=1.6t$. The on-site interaction $U = 0.5t$ (uniform). The subgap states obtained are translationally invariant along the $\hat{\vec{y}}$ direction. Hence, we first construct a projected 1D wavefunction by summing up the components along $\hat{\vec{y}}$ for a given $x$:
\begin{equation}
    \label{SMeq:psi_1D}
    \Psi_{\text{1D}}(x) = \sum_{y}\sum_{l}\left[|u_{l}(x,y)|^{2} + |v_{l}(x,y)|^{2}\right],
\end{equation}
where $u_{l}(x,y)$ and $v_{l}(x,y)$ are the particle and hole components, respectively, of the subgap state at site $\vec{r} = (x,y)$ and orbital $l$. For exponentially localized wavefunctions of the form $\sim e^{-x/\xi}$, the localization length $\xi$ can be estimated as~\cite{kramer1993localization}:
\begin{equation}
    \label{SMeq:loc_length}
    \xi^{-1} \approx \frac{1}{L-1}\sum_{x}\Bigg|\ln \left( \frac{\Psi_{\text{1D}}(x+a)}{\Psi_{\text{1D}}(x)} \right) \Bigg|,
\end{equation}
which is exact in the limit $L \rightarrow \infty$. 

\subsection{$\pi-$junctions in disordered problem}\label{SM:snapshots}
\begin{figure*}
    \centering
    \includegraphics[width=\linewidth]{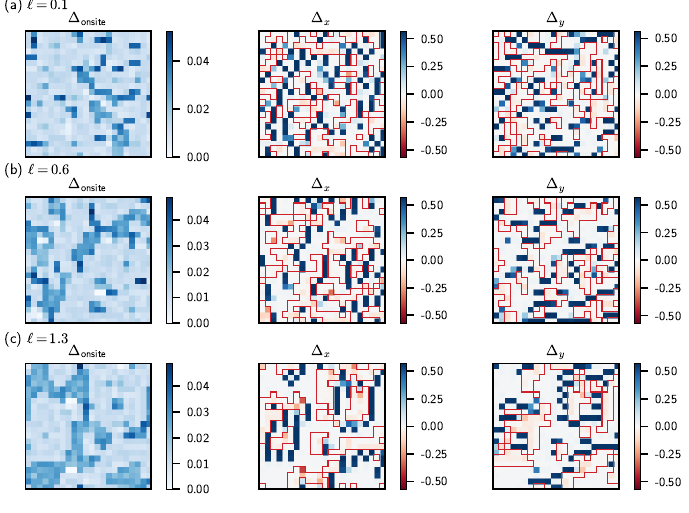}
    \caption{Example of the $s$-wave pairing fields (onsite, $x$ nearest neighbor, $y$ nearest neighbor) solved self-consistently on a $22 \times 22$ lattice. The self-consistent solutions show that $|\Delta_{x,y}|\gg\Delta_{0}$, i.e., the extended components dominate, and that $\Delta_{x,y}$ change sign throughout the system. Results are shown for three different values of the disorder correlation length $\ell$, demonstrating the emergence of larger and larger puddles as $\ell$ grows. The red borders in $\Delta_{x,y}$ mark the boundaries between regions of different sign --- the $\pi-$junctions. Microscopic parameters used in the simulations are $t' = 0.18t, U=0.5t, V_{1}=0.5t, V_{2}=2.1t$, and  fixed $\zeta=0.1$.
    }
    \label{sfig:Delta_snapshots}
\end{figure*}

The attractive nearest-neighbor interaction $V(\vec{r})$ in our model takes two values, $V_1,V_2$ randomly, with the probability distribution introduced in the main text. Since they are both attractive, we expect that the resulting self-consistent pairing fields change sign, thus giving rise to $\pi-$junction physics. In Fig.~\ref{sfig:Delta_snapshots} we show the three pairing fields, $\Delta_0$ (onsite), $\Delta_{x}$ (horizontal nearest neighbor), and $\Delta_{y}$ (vertical nearest neighbor), for a specific disorder realization in a $22\times22$ system, corresponding to three disorder correlation lengths, $\ell$. We find that $|\Delta_{x,y}|\gg\Delta_{0}$, so the dominant pairing is the extended $s$-wave component. Furthermore, our calculation indeed confirms that $\Delta_{x},\Delta_{y}$ change sign throughout the system. The size of the pair field puddles increase with $\ell$, in agreement with the discussion in the main text. 

\subsection{Dependence on disorder averaging}\label{SMsec:disorder_convergence}

\begin{figure*}
    \centering
    \includegraphics[width=\linewidth]{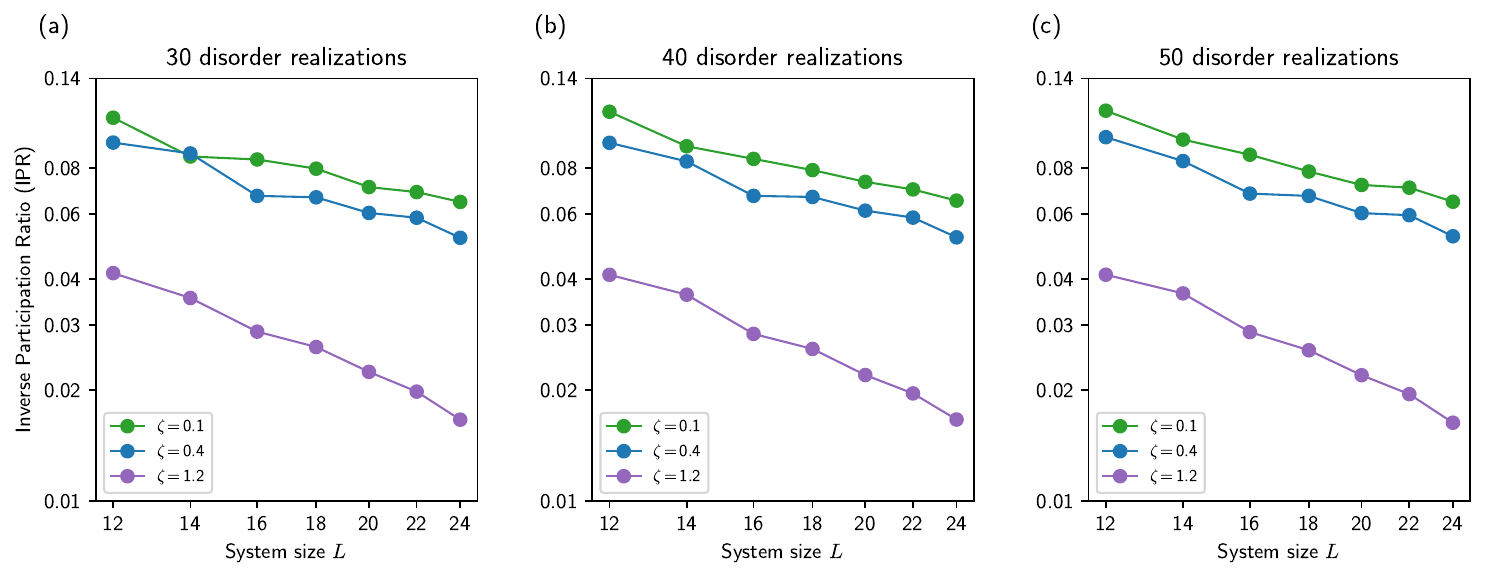}
    \caption{Dependence of the inverse participation ratio scaling with system size on the number of disorder realizations. All other parameters are the same as in Fig.~\figref{fig:general_picture}{c} of the main text.}
    \label{sfig:disorder_window_size}
\end{figure*}

\begin{figure*}
    \centering
    \includegraphics[width=\linewidth]{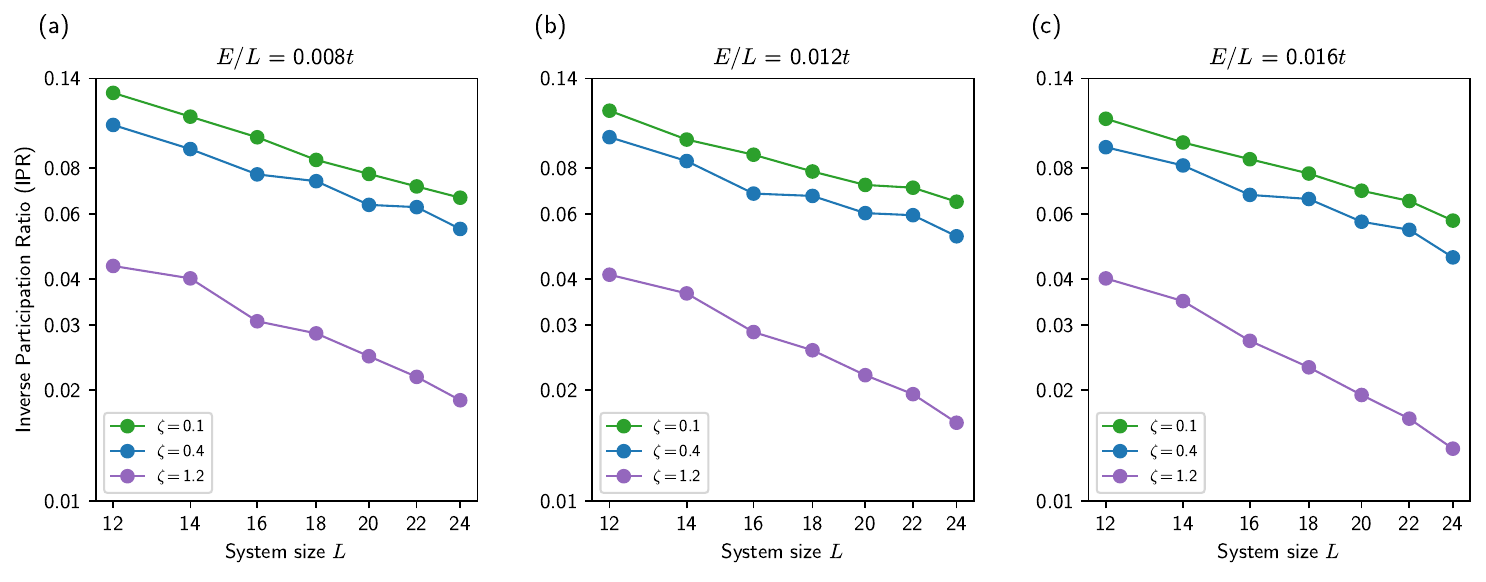}
    \caption{Dependence of the inverse participation ratio scaling with system size on the size of energy intervals $E/L$ over which the IPRs are averaged over. All other parameters are the same as in Fig.~\figref{fig:general_picture}{c} of the main text.}
    \label{sfig:E_L_window_size}
\end{figure*}

In the main text, we have shown results for the IPR averaged over 50 disorder realizations and over a finite energy interval $E/L=0.012t$. Here we demonstrate that our results are already well converged as a function of both of these numbers. In Fig.~\ref{sfig:disorder_window_size} we examine the results averaged over 30, 40, and 50 disorder realizations, and in Fig.~\ref{sfig:E_L_window_size} we show results for subgap states contained in an energy interval $E/L=0.008t,\,0.012t,\,0.016t$. The results converge to the trend we explained, showing that our conclusions do not  depend sensitively on the specific choice of these numbers.

\subsection{Dependence on energy}
\begin{figure}
    \centering
    \includegraphics[width=\linewidth]{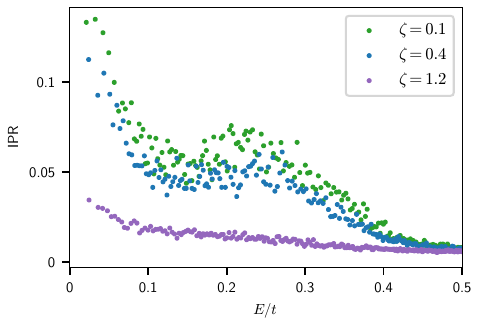}
    \caption{Energy dependence of the inverse participation ratio for a system of linear size $L=24$ and three values of $\zeta$, averaged over 50 disorder realizations. All other parameters are the same as in Fig.~\figref{fig:general_picture}{c} of the main text.}
    \label{sfig:shoulder}
\end{figure}

Figure~\ref{sfig:shoulder} shows the dependence of the IPR on the energy of the subgap states, for a fixed system size $L=24$ and for three values of $\zeta$. For small $\zeta$, we find a ``shoulder": the IPR first decreases with increasing energy, then reveals a plateau region, and then eventually drops off. This behavior disappears at large $\zeta$ and gives way to a monotonic decrease in the IPR as a function of energy. At large enough energies, the states have a large localization lengths regardless of $\zeta$. The energy window we consider when averaging the IPR is below the shoulder threshold, as we are primarily interested in the low-energy properties associated with states that lie far below the clean superconducting gap. A detailed theoretical investigation of the origin of this shoulder feature remains an interesting future direction.

\subsection{Superfluid stiffness calculation}
Introducing the phase twist $\vec{q}=q_{x}\hat{\vec{x}}$ to the Cooper pairs is equivalent to modifying the hopping terms of the Hamiltonian by the Peierls substitution as: 
\begin{align}
h_{\nu}(i,j,\zeta)
&\;\rightarrow\;
e^{ i\frac{\mathbf{q}}{2}\cdot(\mathbf{r}_{i}-\mathbf{r}_{j}) }\,
h_{\nu}(i,j,\zeta),
\nonumber\\
&\qquad\qquad
\nu \in \{1,2\}
\label{SMeq:peierls_flatband}
\\[1ex]
\sum_{\langle i,j \rangle, l, s}
t'\, c_{i,l,s}^{\dagger} c_{j,l,s}
&\;\rightarrow\;
\sum_{\langle i,j \rangle, l, s}
t'\, e^{ i\frac{\mathbf{q}}{2}\cdot(\mathbf{r}_{i}-\mathbf{r}_{j}) }\,
c_{i,l,s}^{\dagger} c_{j,l,s}
\label{SMeq:peierls_dispersive}.
\end{align}
[See Eq.~\eqref{SMeq:zeta-hoppings} for explicit expressions of $h_{\nu}(i,j,\zeta)$].
The added phase twist leads to a change of the free energy $\cal{F}$ due to added superfluid kinetic energy which is proportional to the superfluid stiffness $\rho_{s}$:
\begin{equation}
    \label{SMeq:}
    \rho_{s} = \frac{1}{\Omega}\left. \frac{\partial ^{2}\mathcal{F}}{\partial q_{x}^{2}}\right|_{q_{x} \rightarrow 0}, 
\end{equation}
where $\Omega=L^2$ is the area of the system. Note that we work with the full Hamiltonian with the local interactions (that do not couple to the external vector potential), instead of only an effective Hamiltonian with the interactions projected to the low-energy partially filled bands.  

\section{Role of disorder correlation length}\label{SMsec:ell}
\begin{figure*}
    \centering
\includegraphics[width=0.7\linewidth]{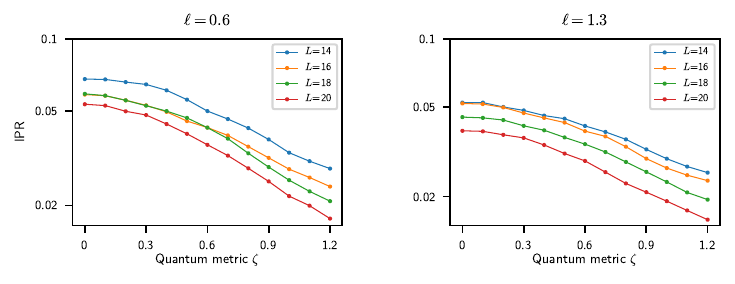}
    \caption{Disorder averaged IPR, averaged over a fixed energy interval $E/L = 0.012t$ plotted as a function of $\zeta$ for several system sizes, for the two disorder correlation lengths $\ell = 0.6$ and $\ell = 1.3$. All other parameters are the same as in Fig.~\figref{fig:general_picture}{c} of the main text.}
\label{sfig:zeta_scans}
\end{figure*}

Apart from increasing the puddle sizes of the pairing fields, changing the disorder correlation length $\ell$ has no qualitative effect on the properties of the subgap states, but affects their localization length quantitatively. To this effect, we perform self-consistent computations of the pairing fields corresponding to two additional disorder correlation lengths in $V(\vec{r})$, for different system sizes and $\zeta$. In Fig.~\ref{sfig:zeta_scans} we show the IPR averaged for an energy interval of $E/L=0.012t$ and over 50 statistically independent disorder realizations. We first notice the drop in IPR of the states at low $\zeta$ compared to those shown in Fig.~\figref{fig:general_picture}{c} of the main text. This is due to the emergence of larger puddles having low (order $10^{-2}t$) positive values of $\Delta_{x}$ and $\Delta_{y}$ in regions where the onsite pairing is relatively large. This results in larger bare coherence lengths $\xi_{0} \sim v_{\rm F}^{\star}/\Delta$ of subgap states located at the $\pi$-junctions in the vicinity of these regions, allowing them to hybridize easily. In addition, we show the disorder-averaged IPR as a function of $\zeta$ for some fixed system sizes. We leave a detailed understanding of the interplay of $\ell,~\zeta$ and the geometry of the domain walls that separate regions of sign-changing gap for future work.

\begin{figure*}
    \centering
\includegraphics[width=\linewidth]{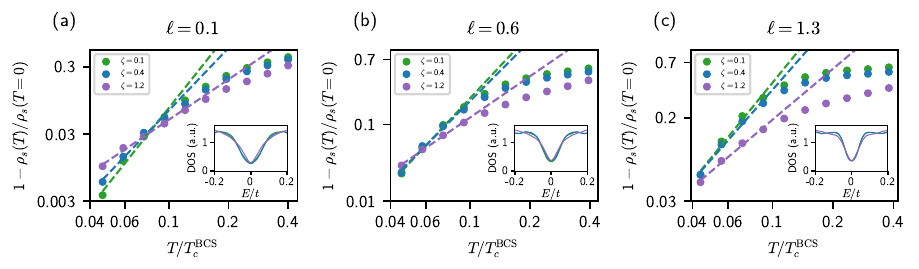}
    \caption{Superfluid stiffness $\rho_{s}$ as a function of temperature $T$ for three disorder correlation lengths (a) $\ell = 0.1$, (b) $\ell = 0.6$, and (c) $\ell = 1.3$, obtained self-consistently in a $18 \times 18$ lattice. Power-law fits (dashed lines) at the low-temperature regime yield exponents which decrease with both $\zeta$ and $\ell$. The stiffness shows a crossover at $T/T_{c}^{\text{BCS}} \approx 0.1$ which corresponds to the low-energy DoS (see inset) transitioning to plateau-like regions. All other parameters are the same as in Fig.~\figref{fig:general_picture}{c} of the main text.}
\label{sfig:stiffness}
\end{figure*}

Figure~\ref{sfig:stiffness} shows the normalized superfluid stiffness in the low- and intermediate-temperature regimes for the three disorder correlation lengths. We perform power-law fits in the low-temperature regimes. The exponents extracted from the fits decrease with increasing $\zeta$ as well as $\ell$. For $\ell = 0.1$, the exponent ranges between $3.7$ at $\zeta = 0.1$, to $2$ at $\zeta = 1.2$; for $\ell = 0.6$, it ranges between $2.8$ to $1.8$, and for $\ell = 1.3$, it ranges between $2.4$ to $1.7$ at the corresponding $\zeta$ values respectively. Interestingly, we observe a crossover in the stiffness-temperature dependence at $T/T_{c}^{\text{BCS}} \approx 0.1$, which is in agreement with the low-energy density of states (see the insets of Fig.~\ref{sfig:stiffness}) showing a crossover to plateau-like regions. This crossover becomes more pronounced with increasing $\ell$. 

\begin{figure}
    \centering
\includegraphics[width=\linewidth]{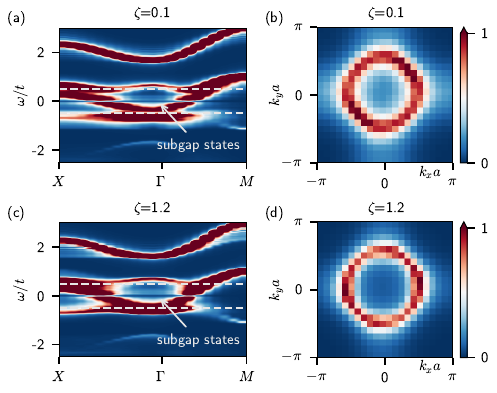}
    \caption{Bogoliubov spectral function ${\cal A}(\vec{k},\omega)$ evaluated on a 22$~\times~$22 lattice (in arbitrary units), for $\zeta=0.1$, $\zeta=1.2$, and disorder correlation length $\ell=1.3$ (same as Fig.~\ref{fig:spectral_function} of the main text in which $\ell=0.1$). (a), (c)~Line cuts of ${\cal A}(\vec{k},\omega)$ along high-symmetry lines in the Brillouin zone. (b), (d)~Momentum-resolved spectral function at zero energy ${\cal A}(\vec{k},\omega=0)$. Other parameters are the same as in Fig.~\figref{fig:general_picture}{c} of the main text.}
\label{sfig:spectral_function}
\end{figure}

\begin{figure*}
    \centering
\includegraphics[width=\linewidth]{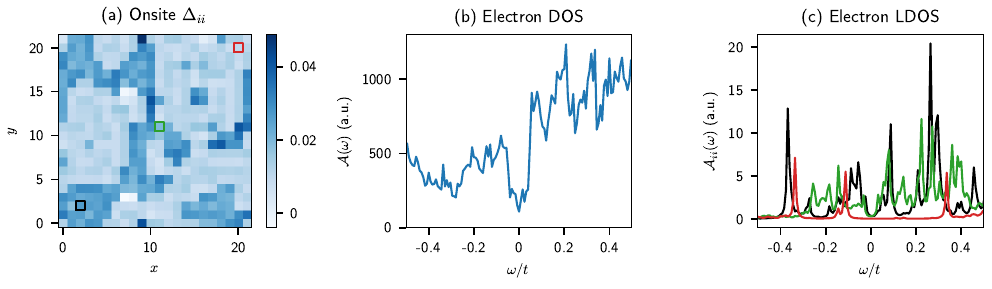}
    \caption{(a)~Onsite pairing spatial map for a specific disorder realization with $\ell=1.3$ [the same as the snapshots in Fig.~\figref{sfig:Delta_snapshots}{c}].
    (b)~Electron density of states as a function of frequency $\omega$, integrated over space. 
    (c)~Electron local density of states, for the three sites marked in black, green, and red in panel (a). The red region, where the extended $s$-wave gap is large, appears locally gapped, whereas the black and green regions, where the extended $s$-wave gap nearly vanishes (and the onsite $s$-wave gap is small) have non-vanishing density of states at low energies. Other parameters are the same as in Fig.~\figref{sfig:Delta_snapshots}{c}.
    }
\label{sfig:LDOS}
\end{figure*}

Figure~\ref{sfig:spectral_function} shows the momentum-resolved spectral function ${\cal A}(\vec{k},\omega)$ for the correlation length of the disorder $\ell=1.3$. These results should be compared with Fig.~\ref{fig:spectral_function} of the main text. For both values of $\ell$ we observe the formation of an impurity ``band" like spectrum and a Bogoliubov ``Fermi surface" of the subgap states. For $\ell=1.3$, the subgap states appear to have a significantly enhanced localization length, consistent with the IPR results shown in Fig.~\ref{sfig:zeta_scans}.
We further analyze the local density of states in Fig.~\ref{sfig:LDOS}, highlighting the difference between regions that are locally ``gapped" and locally gapless regions.

\bibliography{library}

\end{document}